\documentclass[fleqn,10pt]{wlscirep}
\pdfoutput=1  
\usepackage[utf8]{inputenc}
\usepackage[T1]{fontenc}
\usepackage{xcolor}
\usepackage{soul}
\usepackage{cancel}
\usepackage{longtable}
\usepackage{xparse} 
\usepackage{multirow}
\usepackage{graphicx,subcaption}
\usepackage{hyperref}
\usepackage{graphicx}
\usepackage{comment}

\title{Pandemic Vulnerability Index of US Cities: A Hybrid Knowledge-based and Data-driven Approach}

\author[1,*]{Md. Shahinoor Rahman}
\author[2]{Kamal Chandra Paul}
\author[3,8]{Md. Mokhlesur Rahman}
\author[4]{Jim Samuel}
\author[5,9]{Jean-Claude Thill}
\author[6]{Md. Amjad Hossain}
\author[7]{G. G. Md. Nawaz Ali}
\affil[1]{Department of Earth and Environmental Sciences, New Jersey City University, Jersey City, NJ 07305, USA}
\affil[2]{Department of Electrical and Computer Engineering, University of North Carolina at Charlotte, Charlotte, NC 28223, USA}
\affil[3]{The William States Lee College of Engineering, University of North Carolina at Charlotte, 9201 University City Blvd, NC 28223, USA}
\affil[4]{E.J. Bloustein School of Planning \& Public Policy, Rutgers University,  NJ 08901, USA}
\affil[5]{Department of Geography and Earth Sciences, University of North Carolina at Charlotte, 9201 University City Blvd, NC 28223, USA}
\affil[6]{Department of Accounting, Information Systems, and Finance, Emporia State University, Emporia, KS 66801, USA}
\affil[7]{Department of Computer Science and Information Systems, Bradley University, Peoria, IL 61625, USA}
\affil[8]{Department of Urban and Regional Planning, Khulna University of Engineering \& Technology (KUET), Khulna-9203, Bangladesh}
\affil[9] {School of Data Science, University of North Carolina at Charlotte, 9201 University City Blvd, NC 28223, USA}
\affil[*]{mrahman1@njcu.edu}

\begin{abstract}
Cities become mission-critical zones during pandemics and it is vital to develop a better understanding of the factors that are associated with infection levels. The COVID-19 pandemic has impacted many cities severely; however, there is significant variance in its impact across cities. Pandemic infection levels are associated with inherent features of cities (e.g., population size, density, mobility patterns, socioeconomic condition, and health \& environment), which need to be better understood. Intuitively, the infection levels are expected to be higher in big urban agglomerations, but the measurable influence of a specific urban feature is unclear. The present study examines 41 variables and their potential influence on COVID-19 cases and fatalities. The study uses a multi-method approach to study the influence of variables, classified as demographic, socioeconomic, mobility and connectivity, urban form and density, and health and environment dimensions. This study develops an index dubbed the PVI-CI for classifying the pandemic vulnerability levels of cities, grouping them into five vulnerability classes, from very high to very low. Furthermore, clustering and outlier analysis provides insights on the spatial clustering of cities with high and low vulnerability scores. This study provides strategic insights into levels of influence of key variables upon the spread of infections as well as fatalities, along with an objective ranking for the vulnerability of cities. Thus it provides critical wisdom needed for urban healthcare policy and resource management. The pandemic vulnerability index calculation method and the process present a blueprint for the development of similar indices for cities in other countries, leading to a better understanding and improved pandemic management for urban areas and post-pandemic urban planning across the world.   
 
\end{abstract}
\begin{document}

\flushbottom
\maketitle
%
%
\thispagestyle{empty}

\section*{Introduction}
The highly transmittable coronavirus has created a public health crisis that has disrupted many spheres of society across the globe, including health infrastructure, business, economy, education, transportation, and the emotional well-being of people \cite{salama2020coronavirus,samuel2020covid,samuel2020feeling,codagnone2020assessing,ali2021public}. The impacts and the full consequences of the highly contagious COVID-19 infection are still evolving and presently unclear.
While it is established that more than 250 million \cite{worldometers2020covid}
people have been infected across the globe, the factors behind the rapid spread of the COVID-19 infection in urban settings have still not been addressed satisfactorily. At the beginning of this pandemic, many researchers hypothesized that climate and weather could be mitigating factors for the spread of COVID-19 \cite{bashir2020correlation,jamil2020no,rahman2021machine}, but infections in every corner of the world, irrespective of climate and weather, proved the weakness of these and similar conjectures \cite{notari2021temperature}. It has also become part of the dominant narrative that global mega-cities and other metropolitan regions are the zones most severely affected by this pandemic \cite{stier2020covid, shekhar2022leading}. It is commonly understood that cities are the primary hot-spots for COVID-19 infection due to the higher concentration of population and economic activities. Cities in the United States (US) are no exception to the global trend \cite{smith2020covid, rahman2020covid, patel2021covid}. They are said to be highly vulnerable to this pandemic because of their urban vibrancy, including high connectivity, density, socioeconomic makeup, high mobility, and diverse activities with high levels of direct human to human interactions such as personal and cultural services, education and healthcare \cite{cordes2020spatial,sharifi2020covid,nikparvar2021spatio, fortaleza2020taking,rahman2021socioeconomic,deweese2020tale,hanzl2020urban,kashem2021exploring}. Since urban regions house more than 80\% of the US population \cite{bureau_2021} and serve as centers of economic activities, it is important to understand the scale of their vulnerability and their underlying exposure to the pandemic. Furthermore, a clearer understanding of the vulnerability of cities will help us rethink and reshape post-pandemic urban planning strategies and policies. 

It is observed that infection rates are not uniform across cities. While the limiting factors of the spread of COVID-19 in urban areas continue to defy actionable comprehension, many studies  have found relationships between the escalation of coronavirus infections and inherent features of cities (e.g., population size, density, mobility pattern, socioeconomic conditions, health infrastructure, and others). Overall, the infection rate is high in larger urban agglomerations, yet the influence of various urban features remains unsettled. Some researchers have found socioeconomic and demographic indicators to be the pivotal drivers of the pandemic, while others have pointed to urban forms (e.g., employment-household entropy, city size, land-use pattern), mobility patterns as well as transportation systems as significant factors of disease transmission. Fortaleza \textit{et al.} explored the spatial and demographic indicators that affect the vulnerability of populations to the COVID-19 pandemic in 604 municipalities in Brazil\cite{fortaleza2020taking}. They found that the metropolitan land area, population density, how populous a place is, its Human Development Index (HDI), and its Gini index (i.e., income inequality) have a positive association with COVID-19 prevalence \cite{fortaleza2020taking}. Similarly, another study \cite{campos2021vulnerability} from Brazil assessed vulnerabilities of 853 cities and concluded that high urbanization and population density, inadequate sanitation, high illiteracy, high Gini index, low HDI and household income, weak healthcare infrastructure, and a high number of people with underlying health conditions increased urban vulnerabilities to COVID-19 infection. Pourghasemi \textit{et al.} concluded that distance from roads, mosques, hospitals, and density of cities are the main drivers of the diffusion of COVID-19 in Iran \cite{pourghasemi2020assessment}. A study in Uganda found that vulnerability is highest in the areas with shopping malls, transport hubs, and transactional offices \cite{bamweyana2020socio}. Urban areas with more low-income households, weak healthcare infrastructure, poor socioeconomic conditions, and underprivileged communities (e.g., lack of electricity, education, water etc.) are more afflicted by the COVID-19 pandemic \cite{coven2020disparities,coelho2020assessing,imdad2020covid}. Furthermore, a study across 276 Chinese cities observed that urban governance capacity, as measured by the city comprehensive development index, which is based on indicators of environment, society, and economy, is an important factor to control the COVID-19 pandemic, particularly in the later stages \cite {chu2021determines}. The study reported a 2.4\% increase in COVID-19 case recovery for each 1\% increase in the capacity of urban governance.

Besides international studies, some US-based studies have also strived to identify the underlying factors of the faster spread of coronavirus in urban areas \cite{hanzl2020urban, rhodes2021rapid,snyder2020spatial,hatef2020assessing}. Snyder and Parks studied the spatial variation in socio-ecological vulnerability to COVID-19 at the county level in the contiguous US states. They found that southeastern states are highly vulnerable to this pandemic because their population experiences greater social vulnerability \cite{snyder2020spatial}. Cities with a higher Developing Area Deprivation Index (i.e., more disadvantaged neighborhoods: racial minorities, economically deprived strata) are more susceptible to COVID-19 \cite{hatef2020assessing}. Along the same line, other studies \cite{hanzl2020urban,wadhera2020variation,richardson2020presenting,mikami2021risk,aliberti2021fuller} also found that elderly populations, low-income households, racial minorities, and a higher prevalence of underlying health conditions are associated with a higher susceptibility to the COVID-19 pandemic. On the other hand, Cordes and Castro found a negative correlation of white race, educational attainment, and income with the rate of positive COVID-19 cases \cite{cordes2020spatial}. Also noteworthy, the correlation is positive for the black race in New York City. Moreover, a study\cite{karaye2020impact} found a strong positive association of coronavirus cases with a social vulnerability index (SVI).

Vulnerability assessment has been commonly practiced in various contexts, particularly in relation to extreme events. When  urban assessment is contemplated, vulnerability is the condition determined by physical, social, economic, and environmental factors or processes  that increase the susceptibility of a community to the impacts of hazards \cite{international2004living}. As a result, vulnerability is viewed as a multidimensional, differential, and dynamic phenomenon that encompasses physical, social, economic, environmental, and institutional features \cite{birkmann2006measuring}. The assessment of vulnerability to the spread of COVID-19 is important for examining the preconditions and context of cities to reduce health risks and enhance the resilience of humans and societies. The vulnerability score of a community can be determined using its various underlying characteristics as well as the impacts of past events. Both heuristic and data-driven approaches have been used widely to assess the vulnerability of communities \cite{ho2017spatial, papathoma2019vulnerability,chan2019health,amram2020data,marvel2021covid}. For instance, Ho \textit{et al.}\cite{ho2017spatial} calculated the socio-environmental vulnerability to geriatric depression in Hong Kong using odds ratios estimated by a data-driven binomial logistic regression. Chan \textit{et al.}\cite{chan2019health} used a two-stage dimension reduction statistical method to develop a health vulnerability index from nine indicators in the regions of China's Belt and Road Initiative. Like for natural hazards, data-driven approaches have also been used to assess vulnerabilities of cities to the COVID-19 pandemic. Sangiorgio and Parisi \cite{sangiorgio2020multicriteria} assessed the vulnerability of 257 urban districts based on the relative importance of 56 variables, where they relied on a Generalized Reduced Gradient method for model optimization to produce a metric of COVID-19 risk. Snyder and Parks\cite{snyder2020spatial} assessed county-level socio-ecological vulnerability by aggregating the percentile ranks of 18 indicators. Hamidi \textit{et al.}\cite{hamidi2020does} utilized regression-based structural equation modeling where the natural log of COVID-19 infection rate and death rate are taken as outcome variables and urban form, health facilities, and socioeconomic characteristics as predictors of COVID-19 risk in 913 US metropolitan counties. Sarkar and Chouhan\cite{sarkar2021covid} identified four fundamental dimensions of district level vulnerability in India by applying principal component analysis on 16 indicators. Various machine learning methods have been applied in a wide range of human response estimation studies \cite{s2017information, rahman2021machine}. Many data-driven vulnerability assessment studies have utilized machine learning approaches and have statistically generated weights to estimate the importance of various indicators. Weights of different indicators of community vulnerability can also be assigned by experts. For instance, a recent study\cite{marvel2021covid} proceeded this way to develop a pandemic vulnerability index (PVI) of counties in the contiguous US states. Similarly, Smittenaar \textit{et al.}\cite{smittenaar2021covid} developed a COVID-19 community vulnerability index based on relative weights of 40 indicators elicited from subject-matter experts. The Analytic Hierarchy Process (AHP) heuristic was used to demarcate vulnerability levels of urban districts in India \cite{mishra2020covid}. Since the importance of underlying factors of cities to COVID-19 vulnerability is largely unknown and expertise on this subject matter is limited, we argue that a data-driven approach is advantageous for developing a vulnerability index of US cities.  

This study analyzes an extensive set of variables pertaining to demographic and socioeconomic features, urban form, mobility and connectivity, and health and environmental indicators to understand the compound risk factors inherent to different cities. We leverage a hybrid knowledge-based and data-driven approach to comprehend the associations of complex and interlinked relations among various indicators of urban areas to COVID-19 cases and deaths. We focus primarily on the number of COVID-19 infection cases for the main analysis, and then summarize our validation using death prevalence in the discussion section. Although many previous studies have focused on specific US states and vulnerability indices have been developed for counties, as discussed above, cities are the places on the front line of the COVID-19 pandemic. Therefore, this study aims to develop a PVI at the city level --hereafter referred to as PVI-CI-- by investigating 41 influential variables over 3,069 cities in the contiguous US. The outcomes of this research will help us to understand the most influential variables driving the pandemic in urbanized areas. The vulnerability index can also be used to evaluate risks by studying the combination of hazard and exposure components. The findings will be helpful for city-specific policy development and implementation. They may also reveal the role of sustainability measures towards developing resilient cities. The lessons learned from this research may contribute to post-pandemic city planning.

\section*{Results}
\subsection*{Weights and scores}
The influence of variables retained as indicative of  the prevalence of COVID-19 infections in US cities is studied with variable weights derived with the popular F-Classification (FC) feature selection algorithm. Specifically, a total of 41 variables thematically related to demographics, socioeconomic features, urban form \& density, mobility \& connectivity, and health \& environment dimensions  are used and their relative weights are presented in Table \ref{tab:variable_weights_3_methods}. All five vulnerability dimensions have scores in close range, namely between 0.17 and 0.22. As indicated in Table \ref{tab:variable_weights_3_methods}, demographics (0.22) is identified as the most influential dimension for the spread of COVID-19, followed by mobility \& connectivity (0.21) and health \& environment (0.21), socioeconomic (0.19), and then urban form \& density (0.17).

Among the 10 demographic variables, the size of the younger population (0.15) is identified as the most important factor for COVID-19 transmission. Since the younger generation possesses a greater tendency to socialize and interact with others and to do so without COVID-19 safety protocols, and since much of the lower aged population was excluded from early-stage vaccine coverage, they quite likely contributed to the spread of the virus. Some other demographic variables such as household size (0.14), family size (0.14), the share of the non-citizen population (0.12), the population working outside of the principal city (0.11), the size of elderly population (0.11), and the share of the minority population (0.09) also have high contributions to the spread of COVID-19, as articulated in Table \ref{tab:variable_weights_3_methods}. A higher household and family size indicate that people have more crowded living conditions under the same roof, which increases the risk of infection of family members from one infected member after exposure on large contact network. Moreover, a larger household and family size often indicates precarious socioeconomic conditions that may compromise people's ability to withstand this highly infectious disease via established health care services. The non-citizen and minority populations are often involved in informal job sectors such as food delivery and retail businesses which may also increase the chances of their exposure to COVID-19. Additionally, the non-citizen population has limited access to healthcare services, which makes them reluctant to test and seek health care. The population working outside of the principal city contributes to the COVID-19 diffusion, probably indicating the degree of contacts between place-based communities living a certain distance away from each other. Elderly populations are among the critical victims of the COVID-19 pandemic due to their underlying health conditions. Thus, a higher number of elderly residents increases the probability of viral transmission. Interestingly, the total population (0.03) and people living in group quarters (0.06) show a negligible influence on COVID-19 cases.

Among the socioeconomic variables, congested living is the most influential variable, which has high weight (0.17). Congested living conditions limit the ability of city dwellers to maintain adequate social distancing and heighten the risk of community infection. Therefore, it appears to be one of the most critical socioeconomic determinants of the spread of COVID-19. Likewise, education level (0.15), per capita income (0.13), median income (0.11), poverty level (0.09), and low wage population (0.09) also have a major influence on COVID-19 diffusion. Educated people tend to be health-conscious, and therefore a higher share of educated people in a city makes it relatively less vulnerable to the pandemic. A similar argument can be made for the other socioeconomic factors listed. However, working population (0.04), income inequality (0.07), mobile housing (0.08), and unemployment rate (0.08) have weak or no influence on the number of COVID-19 cases, as presented in Table \ref{tab:variable_weights_3_methods}. 

Residential density and population density which measure the state of city compactness have high weights (0.18 and 0.13, respectively); they have a significant influence in determining the vulnerability of a city. The built-up ratio is also an important variable, which receives relatively high weight (0.13). The number of dependent cities (0.13) around a major city is another influential variable among the ten variables related to urban form \& density due to better transport connection and greater interaction opportunities for social, economic and business activities. Additionally, the retail employment density (0.10) has a relatively robust influence on COVID-19 diffusion, as presented in Table \ref{tab:variable_weights_3_methods}. 
On the other hand, the land area of the city (0.04), the employment density (0.09), the population of dependent cities (0.07), the number of households (0.04), and diversity of land uses (0.08) have a limited effect on the spread of COVID-19 cases. 

\begin{table}[ht!]
    \centering
    \caption{Variables and their corresponding weights estimated by the FC method. 
    }
    \begin{tabular}{p{2.1cm}|p{1cm}|p{7cm}|p{1.2cm}|p{1.2cm}}
    \hline
         \textbf{Dimensions}  & \textbf{Scores} & \textbf{Variables} & \textbf{Symbols}& \textbf{Weights} \\
         \hline
         \multirow{10}{*}{Demographic} & \multirow{10}{*}{0.22}  & Total population (+) & $P_t$ & 0.03 \\
                     &  & Size of younger population (+) & $P_{15}$ & 0.15  \\
                     &  & Size of elderly population (+)  & $P_{65}$ & 0.11 \\
                     &  & Group quarter living (+)  & $P_{gq}$ & 0.06 \\
                     &  & Minority population size (+) & $P_m$ & 0.09 \\
                     &  & Non-citizen population size (+) & $P_{nct}$& 0.12 \\
                     &  & Population working in principal city (-) &  $P_{pc}$& 0.05  \\
                     &  & Population working outside of principal city (+) & $P_{opc}$ & 0.11  \\
                     &  & Household size (+) & $H_s$& 0.14 \\
                     &  & Family size (+) & $F_s$& 0.14 \\
         \hline
         \multirow{10}{*}{Socioeconomic  } &  \multirow{10}{*}{0.19}  &  Mobile housing (+) & $H_{mh}$& 0.08  \\
                      &  &  Working population (+) & $P_w$ & 0.04 \\
                      &  &  Income inequality  (+) & $G_i$&0.07\\
                      &  &  Congested living (+) &  $H_r$ & 0.17 \\
                      &  &  Per capita income (-) & $I_{pc}$& 0.13\\
                      &  &  Median income (-) & $I_m$& 0.11 \\
                      &  &  Poverty level (+) & $P_{bp}$& 0.09 \\
                      &  &  Unemployment rate (+) & $U_e$& 0.08 \\
                      &  &  Education level (-) & $E_{hs}$ & 0.15 \\
                      &  &  Low wage population (+) & $P_{lw}$ & 0.09 \\
                    
         \hline
         \multirow{10}{2cm}{Urban form \& density} & \multirow{10}{*}{0.17}  &   City size (area) (+)  & $A_{sk}$ & 0.04 \\
                    &  &   Housing units (+)  & $H_u$ &0.04 \\
                    &  &   Population density (+)  & $D_p$ & 0.13 \\
                    &  &   Built-up ratio (+)  & $Br$& 0.13  \\
                    &  &   Number of dependent cities (+)  & $N_{dc}$& 0.13 \\
                    &  &   Population of dependent cities (+) & $P_{dc}$& 0.07 \\
                    &  &   Residential density (+) & $R_d$& 0.18  \\
                    &  &   Employment density (+)  & $E_d$ &0.09 \\
                    &  &   Retail employment density (+)  & $D_{re}$& 0.10 \\
                    &  &   Diversity of land uses (-)  & $E_he$& 0.08 \\
         \hline
         \multirow{6}{2cm}{Mobility \& connectivity} & \multirow{6}{*}{0.21}  &   Commute to work (+)  & $V_{mt}$& 0.13 \\
                    &  &   Walkability (-)  & $W_a$ & 0.14 \\
                    &  &   Distance to nearest metropolitan city (-)  & $D_{nc}$ &0.25 \\
                    &  &   Travel duration in public transport (+)   & $P_{t30}$& 0.19 \\
                    &  &   Number of airports (+)   & $N_{ap}$ &0.18 \\
                    &  &   Public transport usage (+)  & $T_p$& 0.12\\
         \hline
         \multirow{4}{2cm}{Health \& environment} & \multirow{5}{*}{0.21}  &  Uninsured population (+)   &  $P_{nhi}$& 0.28 \\
                    &  &  Healthcare employment density  (-) & $D_{hw}$ & 0.14\\
                    &  &  Number of hospitals (-) & $H_{10}$ & 0.17\\
                    &  &  Hospital beds (-)  & $N_b$& 0.13 \\
                    &  & Air quality (+)   & $A_{qi}$& 0.27  \\
        \hline
    \end{tabular}
    \label{tab:variable_weights_3_methods}

\end{table}

The results in Table \ref{tab:variable_weights_3_methods} indicate that, among the variables of human mobility and transport connectivity, distance to the nearest metropolitan city (0.25) is the most influential. This is in line with the strong effect of dependent cities, population density, and congested living mentioned earlier.  Metropolitan living and higher density have made social distancing quite challenging, which renders cities closer to a major metropolitan city more vulnerable to the COVID-19 pandemic compared to others. We also note that travel duration in public transport (0.19) and number of airports (0.18) are significant variables for viral propagation owing to limited opportunity for social distancing and fertile ground for exposure to populations from other regions and countries. As for travel duration, we also observe that the increased use of public transportation and higher travel duration to work exert a strong influence on the spread of the virus due to the possibility of infection from other passengers when social distancing is hardly feasible. On the other hand, walkability (0.14) has a comparatively greater role in reducing viral transmission, which indicates that neighborhoods with a higher walkability score have lower rate of COVID-19 transmission. Overall, all of the variables in the mobility \& connectivity dimension have a sturdy impact on the COVID-19 pandemic.  

Among the variables in the health \& environment dimension, uninsured population (0.28) has the higher weight compared to other variables. Thus, this is another critical variable of viral diffusion which, in turn, contributes to a higher PVI score of a city. Air quality is observed to be another influential variable. Extant research\cite{bashir2020correlation, isphording2021pandemic, rahman2021machine} has indicated that air quality improvement due to lockdown-induced mobility reduction has significant impacts on controlling the COVID-19 diffusion. In particular, we observe here a higher weight (0.27) for the air quality index (Table \ref{tab:variable_weights_3_methods}). Similarly, the number of hospitals and of hospital beds have higher weights (0.17 and 0.13, respectively), which underscores the great impact of healthcare infrastructure on controlling the pandemic. As for to the mobility \& connectivity dimension, all variables in the health \& environment dimension have strong effects on COVID-19 transmission.

\subsection*{Pandemic vulnerability index}
\begin{figure}[ht]
\centering
\includegraphics[width=15 cm]{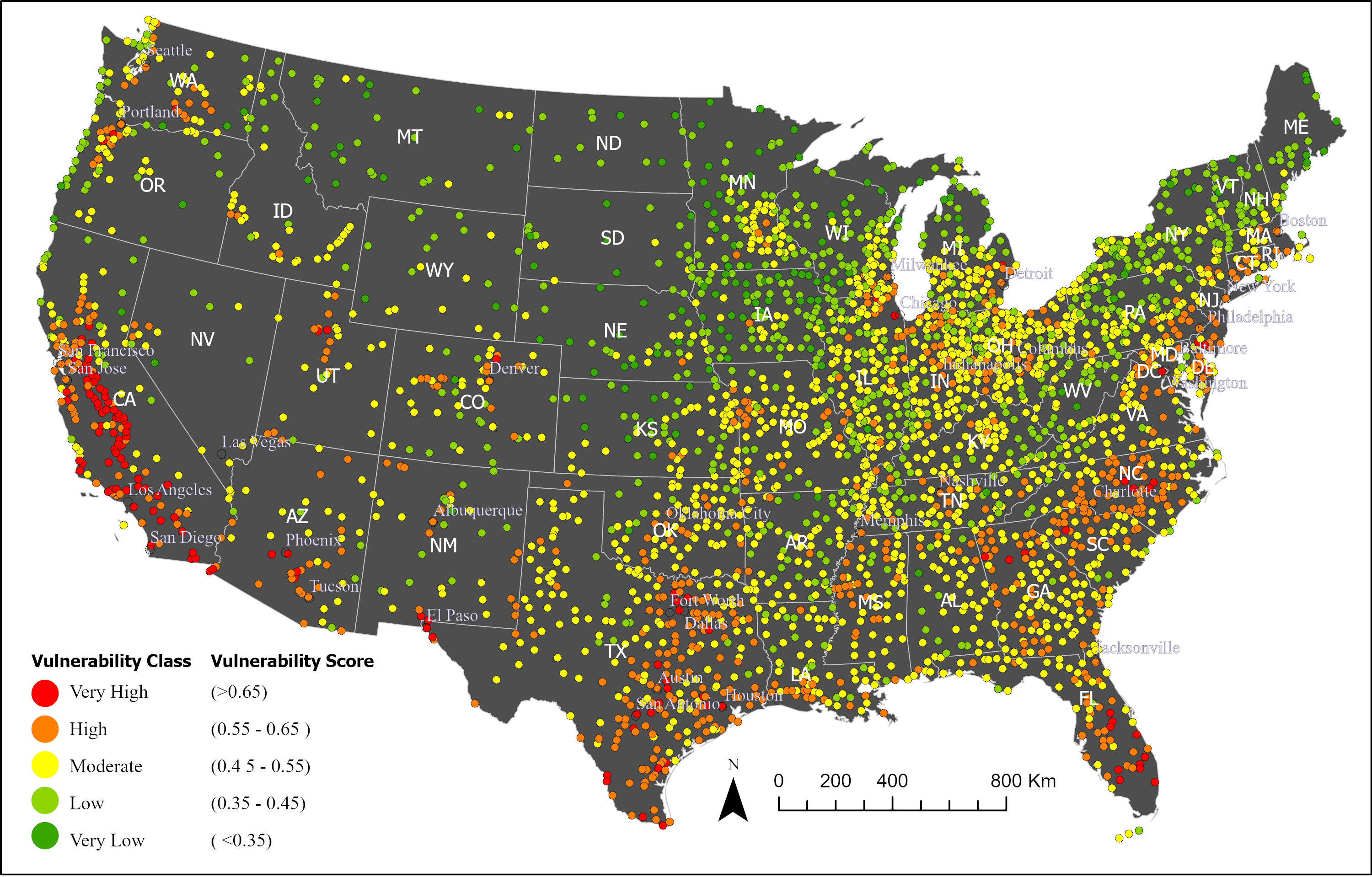}
\caption{Pandemic vulnerability index of US cities.}
\label{fig:Vulnerability_Index}
\end{figure}

Considering the number of COVID-19 cases per 1,000 residents as the target variable, Figure \ref{fig:Vulnerability_Index} portrays city-wise vulnerability categories based on the PVI-CI scores estimated by the FC method. This method was found to provide a better fit than the two other methods considered in this study. The PVI-CI scores are categorized in five classes and visualized for a comparative analysis. Overall, the number of cities with very high, high, moderate, low, and very low vulnerability are 152, 735, 1,390, 682, and 110, respectively. We find that 29\% of US cities have a high or very high level of vulnerability to the COVID-19 pandemic. At the other end of the range, about the same share of cities (26\%) have low or very low vulnerability.   

The map depicts that many cities in California, Oregon, Arizona, Texas, Florida, and North Carolina have a high or very high level of vulnerability to the COVID-19 pandemic. In contrast, cities located in the Rockies (e.g., Montana, Wyoming, New Mexico) and in Midwestern states (e.g., North Dakota, South Dakota, Iowa, Nebraska, Minnesota, Wisconsin) are mostly in the low and very low vulnerability categories. Similarly, many cities located in the Northeast (e.g., Maine, New Hampshire, Vermont, New York, Pennsylvania) face low or very low vulnerability.

\begin{figure}[hb]
\centering
\includegraphics[width=17.5 cm]{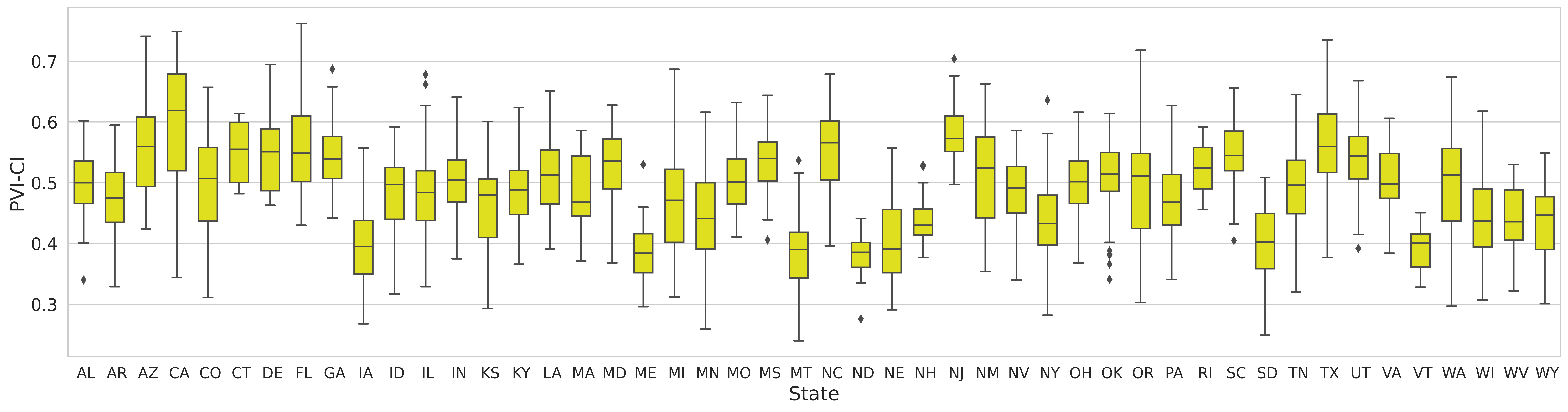}
\caption{Box plots of state-wise PVI-CI scores.}
\label{fig:boxplot}
\end{figure}

Among the 100 most vulnerable US cities, 60 are in California, 14 in Texas, and 9 in Florida. About 70\% of California's 201 cities have either high or very high vulnerability to COVID-19.
However, this is not to say that all cities in California have high vulnerability. A number of them show low vulnerability to the COVID-19 pandemic, such as Portola which is in fact one of the least vulnerable cities in the country. The findings of this study also show that large cities are not always the most exposed to the pandemic, which is a departure from a widely held view. For example, despite being one of the most populous city in the state, San Francisco is not rated as a city with very high vulnerability, like Los Angeles and Riverside, but only features high vulnerability.
Along the same idea, the big cities of New York City, NY, Portland, OR, Denver, CO, Detroit, MI, Charlotte, NC, Philadelphia, PA, and Phoenix, AZ are all highly vulnerable, but other cities in their respective states are worse off and belong to the very high vulnerability category. The case of Seattle, WA, is also noteworthy here. It is one of the largest cities that recorded some of the nation's earliest COVID-19 cases, but it belongs to the high vulnerability category of the PVI-CI. Thus, the size of a city by itself does not determine the level of vulnerability. However, a few large cities are very highly vulnerable. Such is the case of the most populated cities in TX, Houston, Dallas, and San Antonio.

Additionally, we study the distribution of PVI-CI scores by state with a series of box plots (Figure \ref{fig:boxplot}). Each box plot shows the minimum, first quartile, median, third quartile, interquartile range, and maximum of the PVI-CI scores in each state. The median PVI-CI scores in California, Oregon, Arizona, Texas, Florida, and North Carolina are above the highest score of many states. In contrast, the median PVI-CI scores of cities in Montana, Wyoming, New Mexico, North Dakota, South Dakota, Iowa, Nebraska, Minnesota, Wisconsin are below the lowest scores of many other states. Thus, the pattern in PVI-CI scores of cities depicted in Figure \ref{fig:boxplot} is consistent with the findings presented in Figure \ref{fig:Vulnerability_Index}. Pandemic vulnerability shows great heterogeneity across cities, and some of this can be accounted for by differences between states, but not all of it. Some states are also much more homogeneous than others.  

\begin{figure}[ht]
\centering
\includegraphics[width=16 cm]{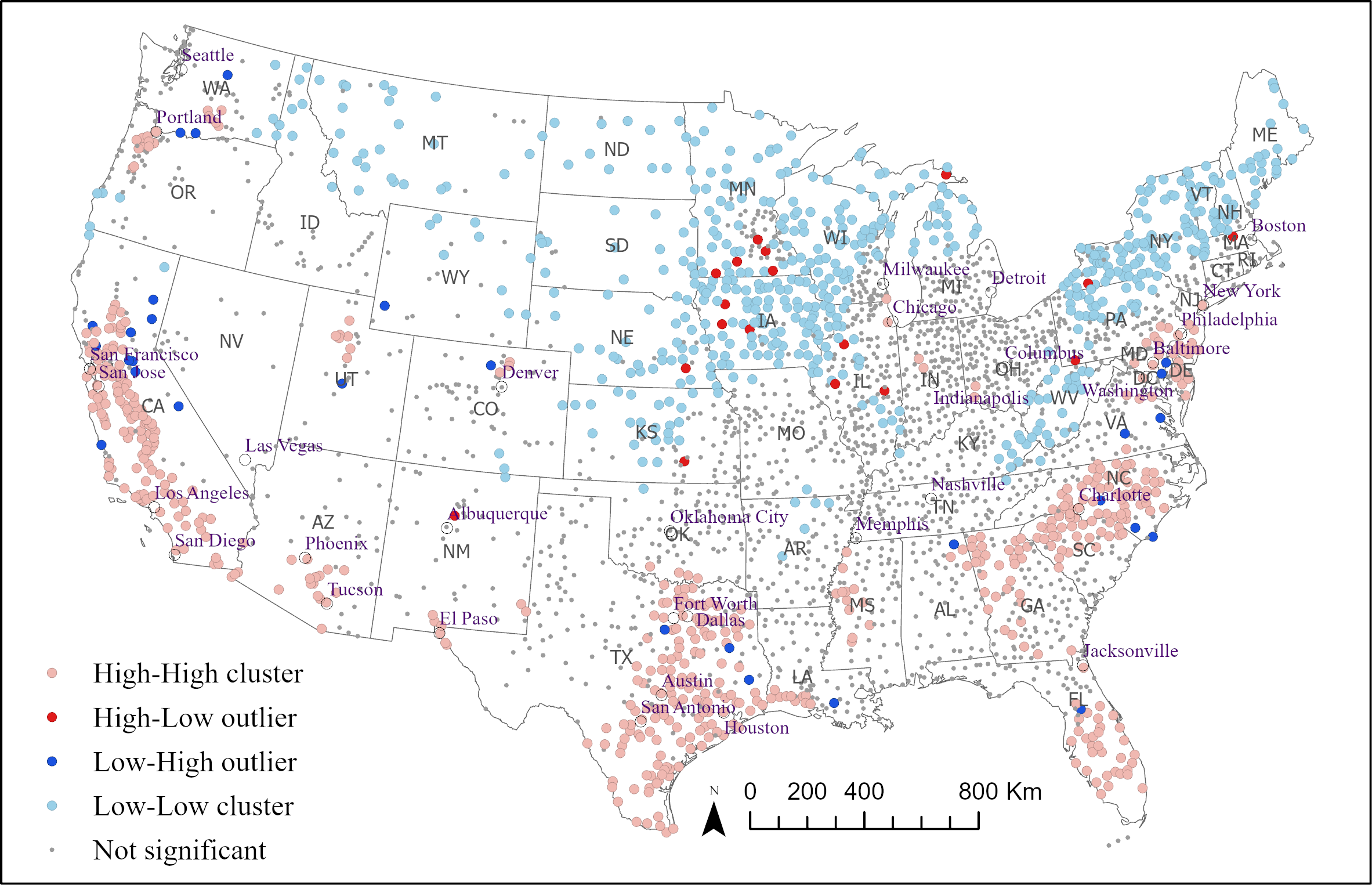}
\caption{Vulnerability clusters and outliers of 3,069 US cities.}
\label{fig:Cluster}
\end{figure}

The mapping of the PVI-CI scores (Figure \ref{fig:Vulnerability_Index}) suggests that vulnerability to the pandemic is not geographically random and that different regions of the country stand out on this basis. Indeed, the global Moran's I of 0.52 indicates that cities are geographically clustered according to their level of vulnerability to the COVID-19 pandemic. To dig further into the nature and properties of this clustering tendency, Figure \ref{fig:Cluster} depicts the results from a local analysis of spatial autocorrelation using Local Moran's I on PVI-CI scores \cite{anselin1995local}. In this analysis, clusters highlight the spatial relationships between cities pertaining to their shared vulnerability characteristics \cite {sanchez2019hot}, while outliers pinpoint the abnormality in the vulnerability behaviors of a city against nearby cities \cite {duan2009cluster}. We find that 35\% of cities are clustered in geographic groupings of similarly extreme vulnerability at the 95\% confidence level. The pink dots in the figure indicate clusters of cities with a higher level of vulnerability (High-High), while light blue dots show clusters of cities of lower vulnerability (Low-Low). On the other hand, red dots denote cities with higher vulnerability surrounded by cities with a lower level of vulnerability (High-Low); dark blue dots represent the reverse situation (Low-High). Cities appearing as gray dots are not statistically significant for being in any cluster or outlier groups.

The geography of higher-vulnerability city clusters (Figure \ref{fig:Cluster}) shows a distinctive pattern. We can delineate five large clusters of higher vulnerability. They encompass California and Southern Arizona, Eastern Texas, South and Central Florida, Northern and Central Georgia and large sections of North and South Carolina, and the Washington DC-Baltimore-Philadelphia-New York City Megalopolis. Some smaller clusters of vulnerability are also found in Northwestern Oregon, Northern Utah, the Front Range Urban Corridor of Colorado, and Western Mississippi. Geographic clusters of lower vulnerability mainly encompass cities at higher latitudes from the Rockies to Upper Michigan and as far South as Kansas and Central Illinois, as well as from New England to the mid-section of Appalachia in Kentucky. 

The outlier analysis shows that a small number of cities behave differently compared to their nearby cities. According to Figure \ref{fig:Cluster}, most of the cities classified as Low-High outliers are located in California, Texas, and North Carolina. For example, Portola in California is a low vulnerability outlier in a High-High cluster due to its small population and low residential density, higher number of children and elderly, low household and family size, lower rate of unemployment, lower number of airports, higher supply of infrastructure per person for health care facilities, high walkability index, lower use of public transport, and better air quality. As discussed earlier, these variables are also regarded as the key factors to contain the vulnerability of cities, despite being spatially located close to cities showing higher vulnerability to the COVID-19 pandemic. The study also shows that, of 18 cities that are High-Low outliers, a preponderance are in three Midwestern states (five in Minnesota, three in Iowa, and three in Illinois). The underlying causes of high city vulnerability to the COVID-19 pandemic discussed earlier are the main reasons for such high-vulnerability outliers (red dots in Figure \ref{fig:Cluster}). 

The study also observes that the ten most vulnerable cities are typically characterized by high population density, large average household size, large average family size, high residential density, high unemployment rate, large share of non-citizen population, long exposure in public transport, low air quality, low per capita income, and low educational attainment. These are the most influential characteristics that make cities vulnerable to the rapid spread of COVID-19. On the other hand, a high share of elderly people, distance from large metropolitan cities, and a large share of working population are associated with low vulnerability. Thus, population and housing density, demographic and socioeconomic characteristics of the population, and the openness of the layout of the city are the salient variables signalling the level of vulnerability of a city to viral infections like COVID-19.

\subsection*{Relative agreement between PVI-CI and NIEHS county level vulnerability}
\begin{figure}[ht]
\centering
\includegraphics[width=16 cm]{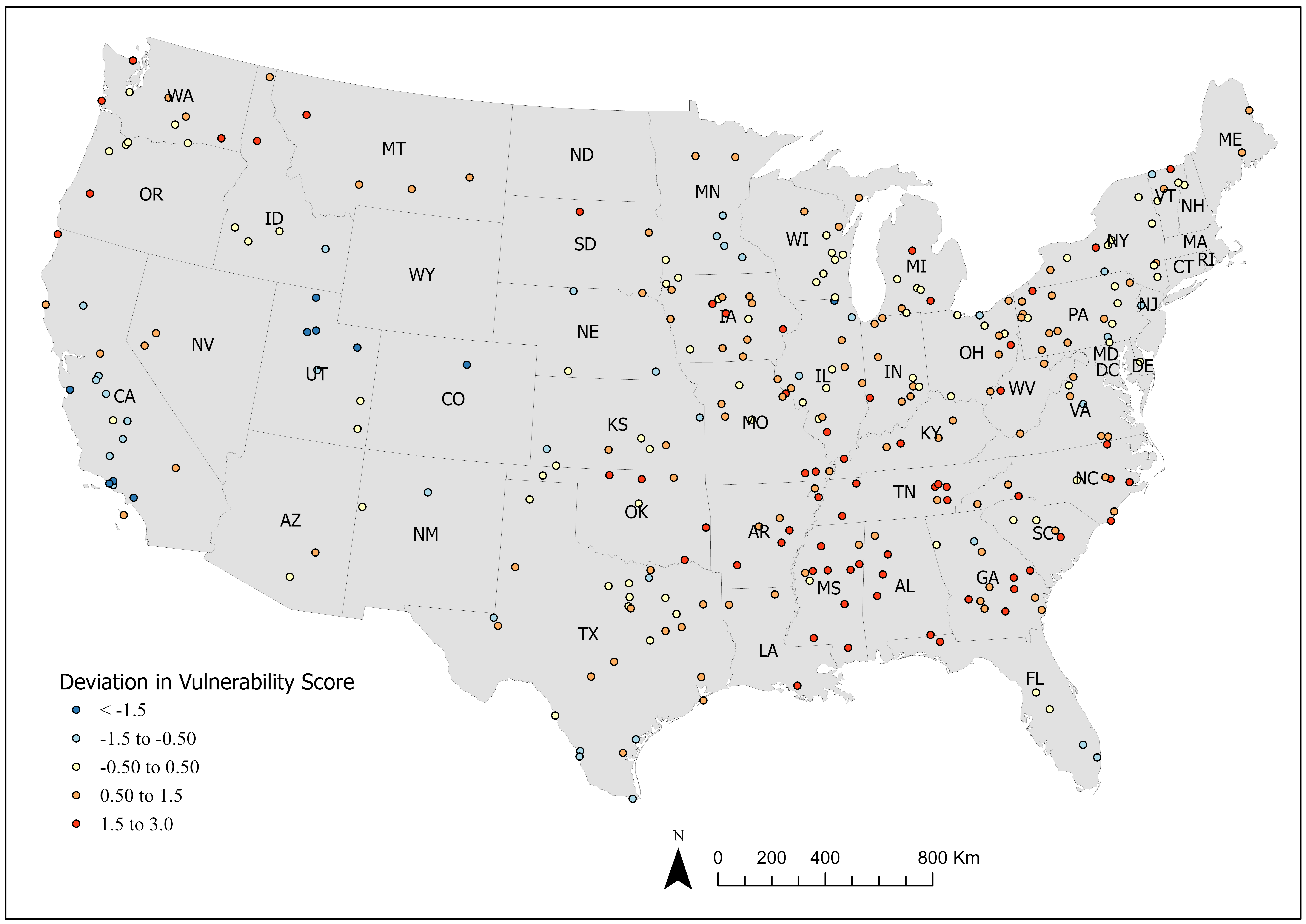}
\caption{Deviation of PVI-CI to the NIEHS vulnerability score for a sample of 300 cities.}
\label{fig:PVI_Deviation}
\end{figure}

We seek to establish the distinctiveness of the PVI-CI by comparing it to the PVI recently devised by the National Institute of Environmental Health Sciences (NIEHS) \cite{marvel2021covid}. We do so for a set of 300 randomly selected cities as the deviation between the PVI score by NIEHS and their corresponding PVI-CI. The Pearson's correlation between the two series is 0.1602, which indicates the two indicators of vulnerability are not equivalent overall, but account for different aspects of the vulnerability of place-based communities to the COVID-19 pandemic. The deviation of standardized scores estimated by NIEHS to their corresponding standardized PVI-CI is mapped in Figure \ref{fig:PVI_Deviation} in five classes around the mean to better contextualize the differences between the two indicators. The vast majority of cities have deviation between -1.5 and 1.5 (75.33\%). In this group, 82 cities have very low positive or negative deviation (-0.5 to 0.5), while a large number of sampled cities (108) have moderately positive deviation between 0.5 to 1.5. Only 36 cities have deviation between -1.5 to -0.5. Specifically, we find that 8 cities have a very low deviation of their PVI-CI score compared to their county NIEHS score (from -0.05 to 0.05), including Jerseyville (IL), Olympia (WA), Laurens (SC), Dumas (TX), Guymon (OK), Winnsboro (SC), Salem (OR), and Guthrie (OK). Also, there are only 10 sampled cities that shows large negative deviation (<-1.5) between NIEHS PVI and PVI-CI scores, against 64 with large position deviations. The 10 cities that show the largest deviations (large positive or large negative) between scores are Tooele (UT), Harvard (IL), Roosevelt (UT), Santa Paula (CA), Longmont (CO), Fairfield Glade (TN), Newport (VT), Sutherlin (OR), Newton (MS), and McComb (MS). The first five cities in this list have higher PVI-CI score compared to the NIEHS score (large negative), whereas the other five are in the reverse situation. A moderately positive skewness in deviations is also observed. Overall, cities with a higher positive deviation in their vulnerability scores tend to be over-represented in the southern states, particularly in Tennessee, Mississippi, Georgia, North Carolina, Arkansas, and Alabama. In contrast, some clusters of cities in the western states (esp. California, Utah, and Colorado) have large negative deviations. However, the large number of cities whose PVI-CI and NIEHS scores are similar (very low deviation) are scattered across the US with no notable spatial trend. It is important to note that there is no discernible variable that is particularly relevant to the deviation between the vulnerability scores. There are some probable data-related reasons for this high deviation between vulnerability scores. First, the cities which have high deviation tend to be located in relatively large counties; thus, these counties are not a good representation of the PVI-CI score at the county level. Second, these cities are often located at the boundary of a county that is under the influence of multiple nearby counties. Third, another city may have the higher influence in that particular county compared to the one which is randomly selected. As a whole, we conclude the PVI-CI is an effective measure of the vulnerability of cities to COVID-19 and this is corroborated by the alignment with the PVI devised by the NIEHS.

\section*{Discussion} 
In this study, we primarily focus on COVID-19 cases to characterize the magnitude, variability and determinants of pandemic vulnerability of cities from the perspective of susceptibility to contract the disease. A related interest calls for the evaluation of pandemic vulnerability from the perspective of mortality due to the disease. The goal here is to see if the vulnerability status of cities is meaningfully different accordingly to the severity of the consequences, that is when we calculate the feature importance based on death rates instead of infection case rates, while following the same methodological steps.

\begin{table}[ht]
    \caption{Result comparison using COVID-19 deaths per 10,000 population as target variable and COVID-19 cases per 1,000 population as target variable.}
    \label{tab:deaths-cases}
    \centering
    \begin{tabular}{|p{2cm}|p{2cm}|p{1.5cm}|p{1.5cm}|p{1.5cm}| p{1.5cm}| p{1.5cm}| p{1.5cm}|}
    \cline{3-8}
         \multicolumn{2}{c|}{}  & \multicolumn{6}{c|}{Number of cities ranked by case counts as target variable} \\ \cline{2-8}
         
         \multicolumn{1}{c|}{} & Class & Very Low & Low & Moderate & High & Very High & Total \\ \cline{1-8}
         \multirow{5}{*} {\parbox{4cm}{Deaths as target \\{variable}}}
         & Very Low & 100 & 10 & 0 & 0 & 0  &  110\\ \cline{2-8}
         & Low & 118 & 496 & 68 & 0 & 0 & 682\\ \cline{2-8}
         & Moderate & 0 & 493 & 815 & 82 & 0 & 1,390 \\ \cline{2-8}
         & High & 0 & 36 & 400 & 299 & 0 & 735\\ \cline{2-8}
         & Very High & 0 & 0 & 21 & 89 & 42 & 152 \\ \hline
        \multicolumn{2}{|c|}{Total} & 218 & 1,035 & 1,304 & 470 & 42 & 3,069\\ \hline

    \end{tabular}
\end{table}
Accordingly, the study also indexes cities with death rates (mortality) as the target variable; the associated results are reported together with results derived on infection cases in table \ref{tab:deaths-cases}. With a focus on death, vulnerability does not precisely align with the conditions for COVID-19 infection and mortality related features are weighted differently from case related features, as demonstrated in Table \ref{tab:deaths-cases}. In other words, this alternative vulnerability index would provide a different ranking for cities by death rates. There could be a number of reasons for this: while the PVI-CI based on cases may point to one ranking for cities, PVI-CI based on deaths can be relatively different because the variables that are associated with deaths due to COVID-19 are different from the variables responsible for the spread of the disease, irrespective of the virulence of the infection. For example, a higher elderly population may raise the death rate because of a host of co-morbidity factors associated with age, but this may not be critical in spreading the disease and increasing the total case count. Both indices based on cases and on deaths can be useful for insights generation and decision making. However, from a policy perspective, by prioritizing the case count as the target variable, governments can improve their understanding of factors that lead to increased case counts, develop policies and plans to mitigate the infection rate, and this can logically be expected to lead to decreased death rates also. 

Results reported in the previous section are based on the use of a single method of analysis of city-wise data, the FC method. As indicated in the next section, two alternative methods have also been considered. The following points can be drawn from a comparison of the three methods. 
\begin{itemize}
  \item While the weights and vulnerability scores were calculated using three alternative methods, the FC method is selected based on the performance metric (i.e., RMSE) to report the results on city vulnerability and discuss the underlying factors of vulnerability. 
  
      \item We also observe some incongruities between the weights provided by the different methods, which cannot be easily reconciled and merits additional research. As an example of a conflicting feature weighting, air quality was ranked as a variable of low influence with the K Best method, while MIR gave it a much higher weight. Similar large differences were observed for the number of hospitals and population of dependent cities.
      
    \item According to the FC method, demographic variables have the greatest influence on city vulnerability, followed by the variables of mobility \& connectivity and health \& environment, urban form \& density, and finally the socioeconomic features of residents. 
    
    \item Population and housing density, demographic and socioeconomic characteristics of the people (e.g., household size, average family size, congested living, share of non-citizen population, per capita income, educational attainment), travel by public transportation, air quality, and the openness of the city are the key considerations to influence the vulnerability of a city.
    
    \item We observe several surprising factors, which include a weak association with the size of the elderly population. This could be due to a greater degree of caution and care among the elderly since the identification of their higher susceptibility levels. Similarly, the total population, which is intuitively anticipated to rank high, displays a relatively weak influence. This could be ascribed to the fact that a large total city population does not necessarily translate into proportionately high population density or proportionately large segments of vulnerable persons.
    
    \item The Cluster and Outlier Analysis shows that cities are strongly clustered geographically based on their vulnerability score (i.e., high vulnerable cities are located in a cluster, while low vulnerable cities are also clustered together). However, some outlier cities are also found. This is the case of low-vulnerability outlier cities located geographically near highly vulnerable cities.
    
    \item The present study evaluates vulnerability scores of cities (PVI-CI) on an aggregate basis using data from a mature phase of the pandemic. It would be quite informative to the pandemic diffusion process to evaluate vulnerability temporally because it can be anticipated that PVI-CI ranked features vary across stages of the pandemic. Therefore, it will also be useful to analyze PVI-CI in a time-phased manner.

    \item The present study will provide an important basis for the understanding of the features associated with city vulnerability to pandemics, epidemic and the spread of contagious diseases more generally. This can help governmental agencies to identify vulnerable spots faster and help advance preparation to mitigate the spread and the mortality rates caused by future variations of COVID-19 or other deadly diseases.  
\end{itemize}

\section*{Methods} \label{sect:methods}
\begin{figure*}[htbp]
\centering
\includegraphics[width=16 cm]{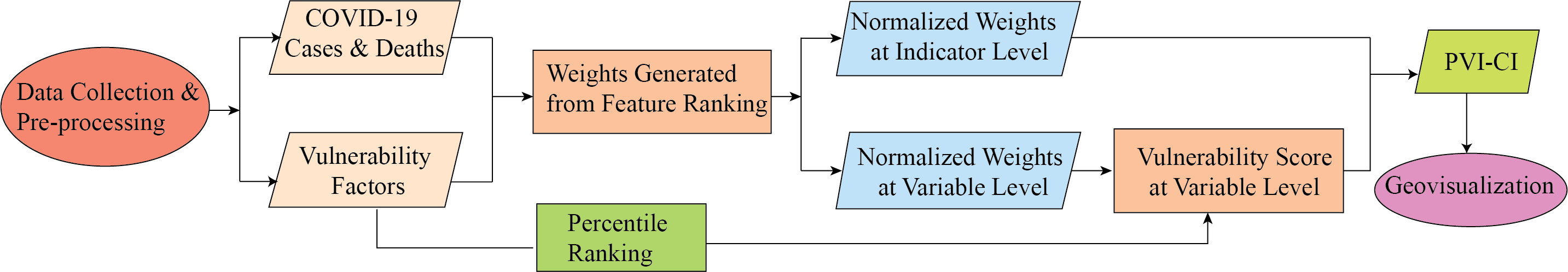}
\caption{Methodological flow diagram.}
\label{fig:flow_diagram}
\end{figure*}

The methodological steps of data collection, cleaning to weight generation, and vulnerability calculation are illustrated in Figure \ref{fig:flow_diagram}. This study utilizes a hybrid approach where weights are determined by a data-driven process and the association of COVID-19 cases with variables are determined by an extensive literature review and expert judgements. At first, based on an extensive review of the literature, a total of 41 important characteristics (factors) of cities are identified. The detailed scope and rationale underlying these factors are discussed in the following subsection. The relative importance of vulnerability factors is then determined through three widely used feature ranking algorithms. Vulnerability scores of cities (PVI-CI) are calculated next on the basis of the weights (importance) of these factors. We end this section with a comparison of the results of the three feature ranking algorithms.  

\subsection*{Data Scoping and Variable Selection}
This study addresses fundamental questions of what drove cities to take the brunt of the COVID-19 pandemic and what the factors of their vulnerability is across the contiguous US. There are over 3,000 cities that vary in size, density, socioeconomic makeup, air quality, healthcare infrastructure, and mobility and connectivity patterns. Thus, this study has collected 41 variables in five dimensions for 3,069 cities to develop a synthetic index of vulnerability to the pandemic. 
The choice of these variables is primarily based on the findings of recent literature related to COVID-19 infection rates and deaths and their association with city environment characteristics. The outcome variables of cumulative COVID-19 cases and deaths were collected on June 30, 2021, from the USA Facts web portal \cite{usafacts.org_2021}. Data on COVID-19 cases and deaths are published for US counties. To obtain city-wise data on the prevalence of the pandemic, county-level data are scaled to the city level using the ratio of city population to the population of the county or counties in which the city is located. 
Tables \ref{tab:demographic_variables} $\sim$ \ref{tab:health_and_environmental_variable} provide detailed description of all the selected variables and their grouping in five broad thematic dimensions.


\subsubsection*{Demographic variables}
Many studies have reported that demographic makeup is one of the prime factors of COVID-19 outbreaks in many locations \cite{azad2021spatio,coven2020disparities,coelho2020assessing,imdad2020covid,kashem2021exploring,xu2021spatial}. A total of 10 variables are selected to apprehend the demographic characteristics of cities (Table \ref{tab:demographic_variables}). Data for all demographic variables are sourced from the urban area census data of the US Census Bureau. All demographic variables are 5-year estimated averages by the American Community Survey (ACS) in 2019, except for the non-citizen population size, which is drawn from ACS's 1-year estimates in 2019.

\begin{table} [ht]
    \centering
    \caption{Description of demographic variables.}
    \begin{tabular}{p{3cm}|p{4cm}|p{9cm}}
        \hline
    \textbf{Factor of pandemic vulnerability} & \textbf{Measurement} & \textbf{Conceptual relevance to pandemic vulnerability} \\
    \hline
    Total population & Total population within the city limits & Total population is an indicator of city size. The larger the city, the greater the opportunities to spread the virus. \\
    \hline
     Size of younger population & Percentage of total population that is under 15 years old & Young population is vulnerable to COVID-19 for two reasons: first, they are mostly not covered by vaccination program; second, they are socially active and less consciously aware of safety protocols.  \\ \hline
    Size of elderly population & Percentage of total population over 65 years old & The elderly population is vulnerable to COVID-19 because of their compromised immune systems and co-morbility conditions.\\ \hline
    Group quarter living & Percentage of total population living in a group living arrangement & People living in group quarter settings are highly exposed to this pandemic since many people live together in a single facility. Thus, the risk of community infection is increased. \\ \hline
   Minority population & Percentage of the minority population in total population & Minority population historically has limited access to education and healthcare services, and less secure income sources, which make them vulnerable to a spreading pandemic.\\ \hline
    Non-citizen population & Percentage of the non-citizen population in total population & Non-citizen population may lack health insurance and legal work status. They are often engaged in informal work, which increases their exposure to a pandemic and reduces their recourse to health care.\\ \hline
    Population working in the principal city & Percentage of working population working in principal city & People working within the city may not travel by public transportation for long duration. They have a lower risk of being infected while commuting. \\ \hline
    Population working outside of principal city & Percentage of working population working outside of principal city & Population working outside the principal city may be at risk because of their long exposure in public transport while commuting.\\ \hline
    Household size & Average number of people in a household & Household size is an indicator of congested living where large households have raised exposure to COVID-19 infection.\\ \hline
    Family size & Average number of people living in a family & Similar to household size, family size is also an indicator of congested living. Larger families have a larger contact network, increasing the risk of COVID-19 infection.\\    \hline
    \end{tabular}
    \label{tab:demographic_variables}
\end{table}

\subsubsection*{Socioeconomic variables}
Fragile socioeconomic conditions make people vulnerable to extreme conditions as well as expose them to the COVID-19 pandemic. As presented in Table \ref{tab:socioeconomic_variables}, a total of 10 social and economic variables are selected in this theme based on their relation to community vulnerability in the extant literature \cite{fortaleza2020taking, campos2021vulnerability, coven2020disparities, snyder2020spatial,kashem2021exploring, hossain2021socio}. The data on all 10 socioeconomic variables are sourced from the US Census Bureau; they are 5-year estimated averages by ACS in 2019.

\begin{table} [ht]
    \centering
    \caption{Description of socioeconomic variables.}
    \begin{tabular}{p{3cm}|p{4cm}|p{9cm}}
        \hline
    \textbf{Factor of pandemic vulnerability} & \textbf{Measurement} & \textbf{Conceptual relevance to pandemic vulnerability} \\
    \hline
    Mobile housing & Percentage of households in mobile homes & People living in mobile housing may face greater economic precarity, which engages them more in informal and essential work activities, where contacts are more common and risk of infection higher. \\ \hline
    
    Working population & Total number of working people & Working population requires to go outside the home daily in the business-as-usual situation.\\ \hline
    
    Income inequality & Gini index of a city& A higher Gini coefficient indicates social disparity, where disadvantaged people may not have adequate access to formal income and avail health care. \\ \hline
    
    Congested living & Percentage of households with 4 or more members & Four or more people in a household indicates a congested living condition, which increases contact across a large network and raises the risk of COVID-19 infection.\\ \hline
    
    Per capita income & Per capita income of workers in a city & Low-income people are most engaged in work with a lot of person-to-person contacts and have high social vulnerability in terms of access to formal work and healthcare services.\\ \hline
    
    Median income & Median household income of a city & Median income of a city indicates financial strength of the city. High median income means city dwellers face lower risk exposure and have higher ability to avoid dire consequences.\\ \hline
    
    Poverty level & Percent of the population below poverty level & People below the poverty level are at high risk of pandemic exposure and lower ability to avoid dire consequences.  \\ \hline
    
    Unemployment rate & Percentage of population 16 years and above with no employment & Unemployed people are often engaged in informal works. hence higher risk exposure and more dire consequences. \\ \hline
    
    Education level & Percentage of the population with high school education and above & Educated people are usually health-conscious, aware of COVID-19 related risks, experience less face-to-face contacts at work, make more use of formal health care, and follow health safety protocols.  \\ \hline
    
    Low wage population & Percentage of the working population paid low wages & Similar to people below the poverty line, low-wage population is more vulnerable to the COVID-19 pandemic. \\ \hline    
    \end{tabular}
    \label{tab:socioeconomic_variables}
\end{table}

\subsubsection*{Urban form and density variables}
Many recent studies have found a relationship between the spread of COVID-19 and variables of urban form \& density \cite{fortaleza2020taking,campos2021vulnerability, pourghasemi2020assessment,xu2021spatial}. This study considers 10 variables to apprehend the important characteristics inherited in cities that may make them hot spots for the COVID-19 infection (Table \ref{tab:ruban_density_variables}). Some variables require further elaboration on how they are constructed. Employment and housing entropy expresses the evenness of the distribution across the types of employment and housing without consideration of the aggregated quantity of jobs or housing \cite{sld2021}. Employment density, residential density, and employment and housing entropy are extracted from the smart location database (SLD) of the US environmental protection agency (EPA) \cite{epa2020}. The values of these density indicators are calculated at the city level by taking the mean of block-group values. The total built-up area of each city is extracted from the GAP/LANDFIRE National Terrestrial Ecosystems database by aggregating developed \& urban land cover classes \cite{mckerrow2014integrating}; the total built-up area is divided by the total city area to get the built-up ratio.   

\begin{table}[ht!]
    \centering
    \caption{Description of urban form and density variables.}
    \begin{tabular}{p{3cm}|p{4cm}|p{10cm}}
        \hline
    \textbf{Factor of pandemic vulnerability} & \textbf{Measurement} & \textbf{Conceptual relevance to pandemic vulnerability} \\
    \hline
     City size (Area) & Total land area & Large cities may have a large population and more economic activities, which increase the risk of disease spreading. \\ \hline

    Housing units & Total number of housing units of a city & Similar to the number of population, the total number of housing units is also an indicator of city size. Many recent studies have pointed that large cities are more exposed to the pandemic.\\  \hline
    
    Population density & Number of population per unit area &  Population density is an indicator of congested living of a city. Higher population density reduces the scope for maintaining social distancing and vice versa. Thus, densely populated areas raise the chance of spreading respiratory diseases.  \\  \hline

    Built-up ratio & Ratio of built-up area to the total area of a city & This is also a density indicator. A high built-up ratio indicates the high density of a city. \\ \hline
    
    Number of dependent cities & Number of cities located within 50 km of a city & This number is an indicator of the spatial setting of the city, if it is the part of a large urban agglomeration versus an isolated city.  \\ \hline
    
    Population of dependent cities & Sum of the total population of cities located within 50km of a city& Similar to the number of dependent cities, people of dependent cities pose an extra risk of disease spreading. \\   \hline
    
    Residential density & Gross residential density (housing units per acre) on unprotected land & Residential density is a density factor directly related to congested living conditions. High residential density lower possibilities for social distancing and is a recognized indicator of higher risk.\\  \hline
    
    Employment density & Gross employment density (jobs per acre) on unprotected land  & High employment density indicates the social interaction of more workers in an area.\\  \hline
    
    Retail employment density &Retail jobs per acre on unprotected land & Retail shops are some of the hot spots of COVID-19 spreading due to face-to-face contacts. Thus, their high density indicates a high risk of virus spreading.\\ \hline
    
    Diversity of land uses & Employment and housing entropy which refers to the mix of employment types and occupied housing& Diversity of land uses indicates that people live mostly near their work places. High land use mix indicates high walkability, which requires a short regular commute. Thus, high entropy of land use mix reduces the COVID-19 infection rate.  \\ \hline
    
    \end{tabular}
    \label{tab:ruban_density_variables}
\end{table}

\subsubsection*{Mobility and connectivity variables}
As indicated in Table \ref{tab:mobility_variables}, a total of six variables are taken into account to assess the mobility and connectivity of cities. Walkability is a component of the built environment \cite{sld2021}. The city-level index is extracted from the National index by taking the mean value of all pixels within the city boundary. Airport locations are derived from the US airports base layer available on the Mapcruzin web portal \cite{mapcruzin}. The distance from each city to the nearest large metropolitan (population 1M+) is calculated by Euclidean distance between city centers. 

\begin{table}[ht!]
    \centering
    \caption{Description of mobility and connectivity variables.}
    \begin{tabular}{p{3cm}|p{4.2cm}|p{8.8cm}}
        \hline
    \textbf{Factor of pandemic vulnerability} & \textbf{Measurement} & \textbf{Conceptual relevance to pandemic vulnerability} \\
    \hline
    Commute to work & Mean vehicle miles travelled per worker per day & Commute to work indicates the length of the trip from home to work. Longer commute may increase the risk of viral exposure.\\  \hline
    
    Walkability & Mean value of the National Walkability Index of all city blocks & High Walkability index implies a large number of walking trips, which reduces travel by public transport. Thus, a city with a high walkability score has a lower risk of disease spreading.\\ \hline
    
    Distance to nearest metropolitan city & Euclidean distance to the city center of the nearest city of one million population  & A city near a large metropolitan city has higher COVID-19 risk compared to a city further away owing to greater interactions. \\ \hline

    Travel duration in public transport & Percentage of population traveling more than 30 minutes on transit & Travel duration in public transport is directly related to COVID-19 infection rate since many people travel together in tight quarters. Long exposure in public transport increases the risk of infection.\\   \hline

    Number of airports & Total number of national and international airports within 50 km from the city boundaries& Airports are the initial hub of disease spreading across the globe. Many people travel together inside a plane for a long time. Moreover, airports are the hub of the interaction of many people.  \\  \hline
    
    Public transport usage & Percent of population commuting to work using public transportation & Cities with a large share of trips made by public transport usually are at high risk.  \\  \hline
    \end{tabular}
    \label{tab:mobility_variables}
\end{table}

\subsubsection*{Health and environment variables}
Health infrastructure is an important consideration to assess COVID-19 vulnerability because many variables are associated with health infrastructure, including testing, vaccination, health awareness building, and treatments \cite{campos2021vulnerability, pourghasemi2020assessment}. Four variables in the health infrastructure dimension are considered for this study (Table \ref{tab:health_and_environmental_variable}). 
People with underlying health conditions, especially respiratory illness (e.g., asthma), have a high risk of COVID-19 infection. Prolonged exposure to air pollution has direct impacts on public health, including asthma, diabetes, and high blood pressure. Many initial studies have shown that air quality may play a vital role in COVID-19 spreading as well as in the mortality rate \cite{bashir2020correlation,isphording2021pandemic,rahman2021machine}. Therefore, this study considers the air quality of cities as an important variable for the COVID-19 vulnerability index. County-level annual air quality index (AQI) data for 2019 is collected from the Air Data Portal EPA since the year immediately before this pandemic \cite{epa}. The median annual air quality index value of the nearest available county is assigned to each city.

\begin{table} [ht!]
    \centering
    \caption{Description of health infrastructure and environmental variables.}
    \begin{tabular}{p{3cm}|p{4cm}|p{9cm}}
        \hline
    \textbf{Factor of pandemic vulnerability} & \textbf{Measurement} & \textbf{Conceptual relevance to pandemic vulnerability} \\
    \hline
     Uninsured population & Percentage of the total population with no health insurance & Uninsured people have limited health care access; thus, they are often reluctant to be COVID tested and take health care services.\\ \hline
    
    Health worker employment density & Health service jobs per acre on unprotected land &  High health employment density is an indicator of the availability of health facilities in an area. \\ \hline
    
    Number of hospitals & Number of hospitals per 10k population &  Number of hospitals in an area is also an indicator of the status of health infrastructure in a city. \\ \hline
    
    Hospital beds & Number of hospitals beds per 10k population & Similar to the number of hospitals, hospital beds are an indication of the health care status of a city.\\  \hline
    
    Air Quality & Mean Air Quality Index (AQI) of county/counties of a city in 2019 & Chronic exposure to air pollution increases respiratory illness. Thus people living in polluted air are immune-compromised; thus, the risk of COVID-19 illness is high. \\ \hline
    \end{tabular}
    \label{tab:health_and_environmental_variable}
\end{table}

\subsection*{Feature Ranking}
\label{sect:feature_ranking}

\begin{figure}[!ht]
    \centering
    \subfloat
    {\includegraphics[height = 12cm, width=8.9cm]{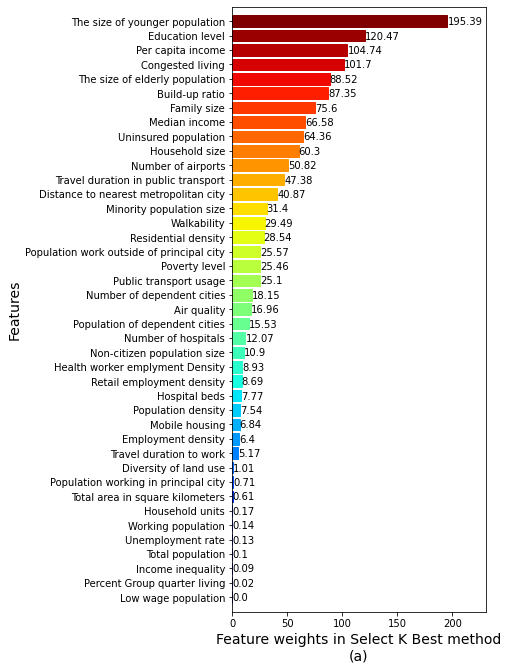}
    \label{Fig:features_skb}
    }
    \subfloat
    {
    \includegraphics[height = 12cm, width=8.9cm]{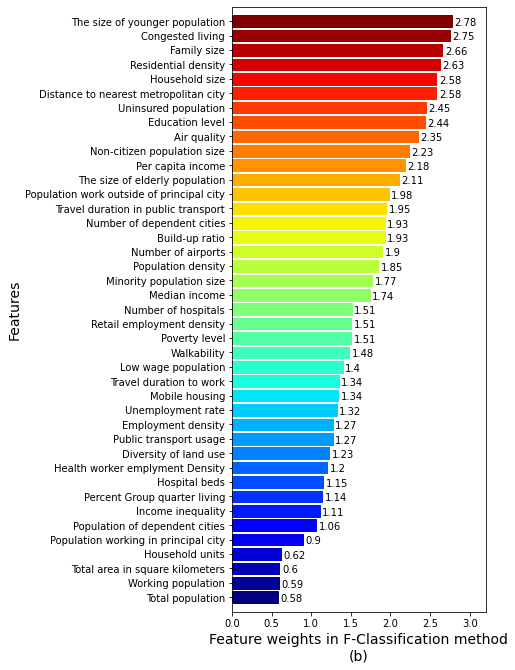}
    \label{Fig:features_fc}}
    \par
    \subfloat
    {
    \includegraphics[height = 12cm, width=8.9cm]{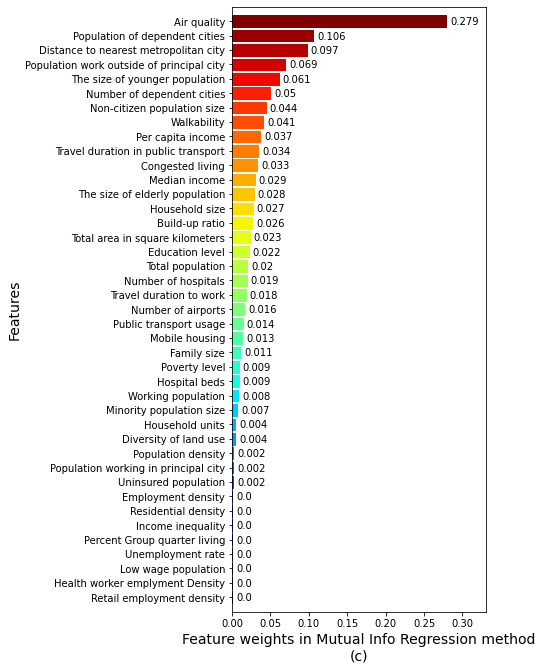}
    \label{Fig:features_mic}}
    
    \caption{Feature ranking using (\textbf{a}) Select K Best, (\textbf{b})  F-Classification, and (\textbf{c}) Mutual Info Regression methods.}
    \label{Fig:Feature_Importance}
\end{figure}

A city-level PVI can be calculated based on a series of factors of the city that contribute to the incidence of COVID-19 cases and deaths. However, not all variables have equal influence on the COVID-19 pandemic. Accordingly, we select a set of 41 most relevant variables to the COVID-19 pandemic, primarily based on the literature review, and thus reflective of expert knowledge and evidence-based research. Then, we use feature ranking methods to rank the selected variables based on their influence on the rates of COVID-19 cases and deaths (Figure \ref{Fig:Feature_Importance}). The ranking of the variables is used to calculate the PVI-CI score of each city.

To strengthen the reliability of the feature ranking, three alternative filter-based methods are used, namely F-Classification, Mutual Information Regression (MIR), and Select K Best (SKB). Each of these methods generate ranking scores of the variables based on different statistical measures. The selected feature ranking methods are presented hereafter.

\begin{enumerate}
    \item \textbf{F-Classification method:} This method selects important features among the variables in a dataset by computing the ANOVA F-value of the sample points \cite{scikit-learn, polat2009new}. Features are often ranked according to their order of importance \cite{costa2018geographical, polat2009new, gunecs2010multi, liang2020machine}. Here, we use the f-classic method of scikit-learn, a Python package, to perform the feature ranking. From the feature ranking, we derive corresponding weights for each feature that serve to determine PVI-CI scores.
    
    \item \textbf{Mutual Information Regression (MIR) method:} In this method, the mutual dependence between the target and an independent variable is estimated based on the mutual information between them \cite {chen2016combining, gao2020relevance, zheng2018feature}. The MIR method incorporates non-parametric algorithms where the entropy is estimated from distances of k-nearest neighbors \cite{kraskov2004estimating, ross2014mutual, doquire2013mutual}. It generates a non-negative value. The larger this value is, the greater the dependency. For two totally  independent variables, the value will be zero. We use the  "mutual\_info\_regression" function of scikit-learn for the implementation \cite{scikit-learn}. The feature weights are calculated from the values of the different features in the ranking.  
    
    \item \textbf{Select K Best Method:} 
    We also use the Select K Best function of scikit-learn to find the ranking of our selected variables based on the k highest scores \cite{scikit-learn, 7300320}. Regression is the scoring function. The derived feature rankings are taken as the weights used towards calculating the city-level PVI, as explained in the following sub-section.
    
\end{enumerate}

\subsection*{Developing the Pandemic Vulnerability Index}
This section describes the methodological steps for determining the Pandemic Vulnerability Index (PVI-CI). After collecting data from secondary sources, we first clean our dataset by removing the cities that have missing values for one or more variables. The final dataset includes 3,069 US cities to assess their vulnerability to the COVID-19 pandemic. Then, we normalize each variable to achieve the same range of values on all features. Since the variables are measured on diverse scales and units (e.g., percentage, absolute number, ratio), normalization brings them in a consistent format which is an essential data prepossessing step for model building. All 41 variables are transformed to the scale between 0 to 1 by max-min normalization. 


The feature ranking methods mentioned above are applied to determine the relative importance of selected variables. These measures serve as the weights ($w$) of corresponding variables in computing the composite vulnerability score of cities. COVID-19 cases and deaths per 1000 population are the target variables (dependent variables) against all other parameters (independent variables). We repeat the following mathematical procedures with each method and determine the pandemic vulnerability index of cities.

 The variable importance scores derived by feature ranking are normalized among the variables of each dimension to determine their corresponding weights ($w$). For instance, the importance scores of the 10 demographic variables are normalized using maximum and minimum values of those variables. In order to find the relative score, cities are sorted either in ascending (+) or descending (-) order for each variable based on their positive or negative association with COVID-19 spreading (sign indicated in Table \ref{tab:variable_weights_3_methods}). Then, the percentile ranks of all cities are calculated for each variable based on the ascending or descending orders as indicated in the aforementioned figure. The vulnerability scores of each dimension is computed as the sum of products of the percentile ranks and of corresponding weights of variables of this dimension, divided by the sum of weights. The vulnerability scores of dimensions are calculated by combining percentile ranks and weights derived from the data driven process using
 equation (\ref{eqn:score_demographic}).
 The vulnerability score for the demographic dimension ($S_{\mathbb{D}}$) is computed as 

\begin{equation}\label{eqn:score_demographic}
    S_{\mathbb{D}} =\frac{ \sum _{d\in \mathbb{D}} (d \times w_d) } 
   {\sum_{ d\in \mathbb{D}} w_d}
\end{equation}

where $\mathbb{D}=\{P_{t},P_{15},P_{65}, P_{gq},P_{m}, P_{nct}, P_{pc}, P_{opc}, H_{s}, F_{s}\}$ and $w_d$ denotes the weight of the corresponding variable $d \in \mathbb{D}$, which is calculated by each feature selection method. In a similar fashion, we find the scores for the variables under the socioeconomic dimension ($S_{\mathbb{S}}$), the urban form and density dimension 
($S_{\mathbb{U}}$), the mobility and connectivity dimension ($S_{\mathbb{M}}$), and the health infrastructure and environment dimension ($S_{\mathbb{H}}$).

The weight of each of the five dimensions is the mean value of feature importance scores derived with the feature ranking methods using equation \ref{eqn:avg_weights}: 

\begin{equation} \label{eqn:avg_weights}
    W_{avg}(\Delta)= \frac{\sum_{k \in \Delta} w_k}{|\Delta|}
\end{equation}

where $\Delta \in \{\mathbb{D, S, U, M, H}\} $. Here, $\mathbb{D}$, $\mathbb{S}$, $\mathbb{U}$, $\mathbb{M}$, and $\mathbb{H}$  represent demographic, socioeconomic, urban form and density, mobility and connectivity, and health and environment dimensions, respectively. The average weight $W_{avg}(\Delta)$ of each dimension is normalized using equation \ref{eqn:dem_norm_weight}: 

\begin{equation} \label{eqn:dem_norm_weight}
    W_{norm}(\Delta) = \frac{ W_{avg}(\Delta)}{\sum_{\Delta \in \mathbb{\{D, S, M, U, H}\}} W_{avg}(\Delta)}
\end{equation}

where $W_{norm}(\Delta)$ denotes the normalized weight of the corresponding dimension. Finally, a city's PVI-CI score is derived from the percentile ranking and normalized weights of different dimensions using equation \ref{eqn:pvi_score_of_city}: 



\begin{equation} \label{eqn:pvi_score_of_city}
    PVI-CI =
    \sum_{\Delta \in  \mathbb{\{D, S, M, U, H}\}}(
    S_{\Delta} \times W_{norm}(\Delta))
\end{equation}

where $S_{\Delta}$ denotes the scores of dimensions for that city. 

The PVI-CI scores of cities are visualized in Figure \ref{fig:Vulnerability_Index} using a choropleth mapping technique available in the ArcGIS Pro software. The figure clearly shows the geographic pattern in the variation of the pandemic vulnerability and its statistical range from very high vulnerability to very low vulnerability. We further compared the scores derived in our study with the scores calculated by the NIEHS for a 10\% random sample of cities. Although NIEHS calculated their pandemic vulnerability index at the county level, we have compared the PVI-CI scores with them to gauge the level of consistency. Moreover, there are no other studies in the literature that would have estimated the vulnerability of cities under the spread of the COVID-19 pandemic. Thus, PVI-CI scores from the most effective of the three feature ranking methods are compared with the PVI scores provided by NIEHS \cite{marvel2021covid}.

After determining the vulnerability scores for all 3,069 cities using the most effective method, a cluster and outlier analysis is performed with the Anselin Local Moran's I to identify the spatial clusters of cities with high and low vulnerabilities. Cluster types for each statistically significant city are determined based on Local Moran's I value and z-score at the 95\% confidence level. A positive Moran's I statistic indicates a city has neighboring cities with similarly high or low vulnerability scores. In contrast, a negative Moran's I indicates that a city is surrounded by cities with dissimilar vulnerability levels \cite {sanchez2019hot}.

\subsection*{Comparison of Feature Ranking Methods}
It is impossible to make a strict validation of the outcomes of this research because of the unavailability of ground truth for validation. Therefore, this study provides an indicative soft validation by comparing the PVI-CI score with the county-level vulnerability score developed by the NIEHS \cite{marvel2021covid}. Thus, technically, this process can be called a relative comparison with an available and established vulnerability index instead of a strict validation. The main goal of this comparison is to select the feature ranking method that performs relatively better to calculate vulnerability scores for all cities in the US. In addition, this comparison process can tell us how close the vulnerability score of a city is related to their corresponding county. 

\begin{figure}[ht]
    \centering
    \subfloat
    {\includegraphics[width=5.5cm]{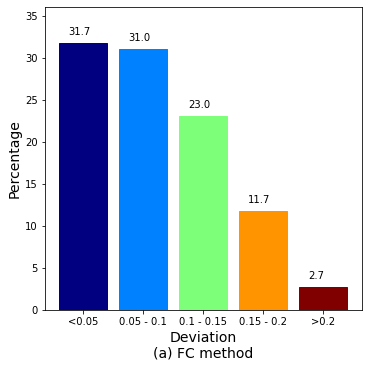}
    \label{Fig:valid_skb}
    }
    \subfloat
    {
    \includegraphics[width=5.5cm]{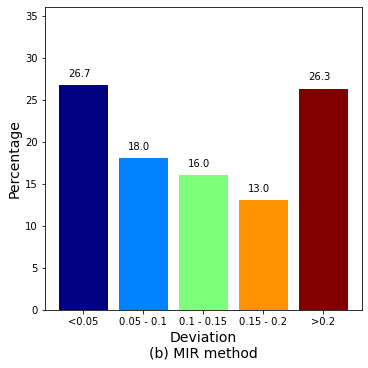}
    \label{Fig:valid_fc}}
    \subfloat
    {
    \includegraphics[width=5.5cm]{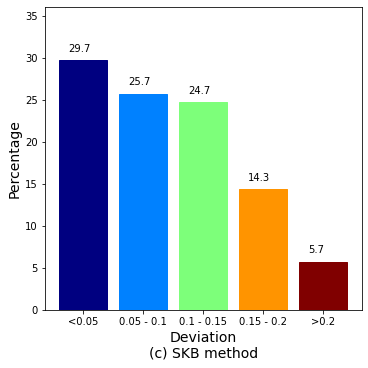}
    \label{Fig:valid_mic}}
    
    \caption{Performance comparison of PVI-CI with respect to NIEHS county level PVI for (\textbf{a}) F-Classification, (\textbf{b}) Mutual Information Regression, and (\textbf{c}) Select K Best  methods.}
    \label{Fig:validation_NIEHS}
\end{figure}

A total of 300 cities, about 10\% of the 3,069 cities, are randomly selected for validation purposes. The PVI scores of 300 counties corresponding to the sampled cities were collected from the PVI dashboard of the NIEHS. Although the PVI scores by the NIEHS are at the county level, this can be utilized as proxy ground truth for the city score since PVI scores at the city level are not available. We have used county-level PVI data to represent scores of cities. In most cases, the representative county lies within the city. Only a few cities in the selected sample lie in more than one county. In that case, the PVI score of a county consisting of the city center is taken into account. At first, the absolute differences between the PVI scores of the NIEHS and our individual proposed methods are calculated for each of the 300 cities. Then, the sum of the squared differences is calculated and divided by the number of sampled cities (300). Finally, the root means square error (RMSE) of the PVI scores for the selected cities are calculated by taking the square root of the above value. This procedure is repeated for all three feature ranking methods (FC, MIR, and SKB). From the deviations of the PVI-CI scores with respect to the NIEHS method, we have also calculated the number of cities that lie within the deviation range of 0.1. The purpose of this calculation is to see how many cities have PVI-CI scores closer to the NIEHS PVI scores. The deviation in vulnerability scores of the sampled cities are presented in Figures \ref{Fig:validation_NIEHS} and \ref{fig:PVI_Deviation}.  Both PVI-CI and NEIHS scores are normalized using Z-score normalization to calculate the deviation between corresponding scores. A geo-visualization of the deviation between pairs of scores is presented in Figure \ref{fig:PVI_Deviation}.

Figure \ref{Fig:validation_NIEHS} depicts the deviation of PVI scores of sampled cities calculated by alternative feature ranking methods from the respective county's scores determined by the NIEHS. The results show that the FC method yields the lowest deviation in scores. A total of 31.7\%, 24.7\%, and 29.2\% of sampled cities have PVI score deviation below 0.05 with the FC, MIR, and SKB methods, respectively. Cumulatively, 62.7\%, 44.7\%, and 55.4\% of sampled cities have PVI score deviation less than 0.10 with these three methods, respectively. In contrast, more than one-third of sampled cities have a PVI score deviation greater than 0.15 with the MIR method. Based on this evaluation, it can be concluded that both FC method and SKB generates vulnerability scores of sampled cities that are the closest to the PVI scores from the NIEHS. In addition, we have also calculated the RMSE to select the most effective method based on how close the PVI-CI score is to the PVI of their corresponding counties. The RMSEs of the FC, MIR and SKB methods are 0.097, 0.155, and 0.111, respectively. As a result, the FC method can be considered as the best one to determine the vulnerability scores for cities.

\section*{Outcomes}

Based on this study, we find the following significant outcomes:  
\begin{enumerate}
    \item City size (population or area) has limited influence on pandemic vulnerability. It is not evident that big cities always have a very high level of vulnerability to the pandemic. For instance, San Francisco, New York, Chicago, and Philadelphia do not lie in the very high vulnerability group, although they are considered as some the largest in the country. 
    \item Cities in California, Texas, Florida, Arizona, and North Carolina are mostly in high and very high vulnerable categories.
    \item The most influential urban vulnerability demographic variables are the size of the younger population and family size. Congested living and educational attainment are influential factors along the socioeconomic dimension. Retail employment density is a highly weighted variable among all variables in the urban form and density dimension. Distance to the nearest big metropolitan city is the most significant mobility and connectivity variable. Both uninsured population and air quality are also determinant variables for making cities vulnerable to COVID-19 spreading.
    \item Although county-level vulnerability and city-level vulnerability are not consistently related, city-level vulnerability scores are close to the vulnerability of corresponding counties. 
\end{enumerate}

\section*{Conclusions}
This research makes significant contributions on three different levels: a) it factors a broad range of variables to identify the pandemic vulnerability ranking of US cities; b) it demonstrates the critical factors of pandemic vulnerability using a multi-method approach, which generates deeper insights in the form of variations in weights of the variables, with an evaluation of case and mortality rates as target variables and detailed interpretation using case rates; and c) it points out dimensions that require further attention and vulnerability modeling, namely the need to study temporal aspects of vulnerability, conflicting weights for a few important variables indicating possible interactions between variables, and the need for better integrated and more detailed data for future PVI modeling. This study is therefore expected to be valuable to governmental agencies in their effort to manage pandemic-associated variables, as well as for future PVI and urban vulnerability research and post-pandemic urban planning. 

\section*{Data availability statement}
The datasets used and/or analyzed during the current study available from the corresponding author on reasonable request.

\bibliography{Bibliography}



\section*{Author contributions statement}

This work was completed with contributions from all the authors; M.S.R conceived the experiments, M.S.R., K.C.P., and M.M.R collected and curated data; M.S.R and K.C.P conducted the experiment; M.S.R, K.C.P., M.M.R., and M.A.H. analysed the results; validation, M.M.R., K.C.P., and M.S.R.; writing—original draft preparation, M.S.R., M.M.R., G.G.M.N.A.,  M.A.H.,  K.C.P., J.S., and J.-C.T.; writing—review and editing, M.S.R., G.G.M.N.A., M.M.R., K.C.P., J.S. and J.-C.T.;  funding acquisition, J.-C.T. All authors edited, reviewed and improved the manuscript. All authors have read and agreed to the published version of the manuscript.

\section*{Appendices}

\begin{longtable}{p{6.2cm}|p{1.35cm}|p{1.5cm}|p{1.65cm}|p{1.5cm}|p{1.5cm}} 
\caption{Descriptive statistics of the variables used in developing PVI-CI (N = 3,069).} \label{tab:variable_statistics} \\
\hline
\textbf{Description of Variables} & \textbf{Measure}     & \textbf{Mean} & \textbf{Std. Dev.} & \textbf{Minimum} & \textbf{Maximum} \\
\hline

\multicolumn{6}{l}{\textbf{COVID-19 Pandemic}}                                                                                                                                               \\ \hline
Number of COVID-19 Cases per 1,000 people (through July 2021)                                                  & \#                   & 8,267.43       & 61,252.30           & 0.00             & 2,221,865.00       \\
Number of COVID-19 Deaths  per 1,000 people (through July 2021)                                                 & \#                   & 147.75        & 1,363.07            & 0.00             & 60,409.00         \\ \hline

\multicolumn{6}{l}{\textbf{Demographic variables}} \\ \hline
Total population                                                   & \#                   & 81,920.78      & 556,295.34          & 1,263.00          & 18,750,806.00      \\
Percentage of population 15 years and below                                             & \%                   & 19.20         & 4.50               & 2.20             & 43.10            \\
Percentage of population 65 years and above                                             & \%                   & 17.31         & 5.58               & 0.00             & 65.10            \\
Average number of people in a household                                           & \#                   & 2.51          & 0.39               & 1.71             & 8.08             \\
Average number of people living in a family                                              & \#                   & 3.14          & 0.38               & 2.00             & 7.94             \\
Percentage of population living in a group living arrangement                           & \%                   & 5.58          & 8.33               & 0.00             & 72.95        \\
Percentage  of non-citizen population of total population                              & \%                   & 3.98          & 5.60               & 0.00             & 56.59            \\
Percentage of population working inside of principal city                          & \%                   & 7.84          & 21.61              & 0.00             & 100           \\
Percentage of population working outside of principal city                         & \%                   & 39.24         & 46.14              & 0.00             & 1001           \\
Percentage of minority population                                   & \%                   & 19.88         & 19.00              & 0.00             & 99.80            \\ \hline

\multicolumn{6}{l}{\textbf{Socioeconomic variables}}                                                                                                                                         \\ \hline
Percentage of mobile homes and others                                                   & \%                   & 7.36          & 7.80               & 0.00             & 86.70            \\
Total number of working people                                                             & \#                   & 37,151.08      & 257,723.98          & 373.00           & 8,658,236.00       \\
Gini index of a city                                                                  & Gini ratio           & 0.44          & 0.05               & 0.26             & 0.77             \\
Percentage of households with 4 or more members                                        & \%                   & 20.94         & 7.56               & 2.10             & 78.80            \\
Per capita income of workers                                                & \$                   & 25,386.36      & 7,364.63            & 7,226.00          & 74,970.00         \\
Median household income                                                    & \$                   & 61,346.18      & 18,131.14           & 0.00             & 173,565.00        \\
Percentage of population below poverty level                                            & \%                   & 18.32         & 8.76               & 0.92             & 55.40            \\
Percentage of population 16 years and above with unemployment                           & \%                   & 6.18          & 3.68               & 0.00             & 39.80            \\
Percentage of population with high school education and above                           & \%                   & 85.35         & 8.88               & 22.00            & 100.00           \\
Percent of total working population who is paid low wage, 2017                       & \%                   & 0.14          & 0.07               & 0.00             & 0.42             \\  \hline

\multicolumn{6}{l}{\textbf{Urban form and density variables}}                                                                                                                                \\ \hline
Total land area, 2018                                                              & km2                  & 87.37         & 391.16             & 0.74             & 9,469.07          \\
Total number of housing units, 2018                                                 & \#                   & 33,732.10      & 218,242.43          & 497.00           & 7,441,836.00       \\
Population density, 2018                                                                         & \#/km2               & 679.88        & 301.75             & 151.31           & 3,597.28          \\
Ratio of built-up area to the total area, 2018                                   & Ratio                & 0.46          & 0.16               & 0.00             & 0.92             \\
Number of cities within 50km of a city, 2018                                           & \#                   & 9.56          & 5.71               & 1.00             & 51.00            \\
Population of cities within 50km of a city, 2018                             & \#                   & 544,644.46     & 1,126,379.56         & 2,198.00          & 24,684,067.00      \\
Residential density, 2018                                                                      & Unit/Acre           & 0.63          & 0.69               & 0.00             & 23.38            \\
Employment density, 2017                                                                       & Job/Acre            & 0.65          & 0.65               & 0.00             & 10.14            \\
Retail employment density, 2017                                                                   & Job/Acre            & 0.08          & 0.08               & 0.00             & 1.15            \\
Diversity of land uses, 2017                                                                      & Ratio                & 0.29          & 0.13               & 0.00             & 0.74      \\  \hline

\multicolumn{6}{l}{\textbf{Mobility and connectivity variables}}                                                                                                                             \\ \hline
Mean Vehicle miles travelled per worker per day, 2020                                              & Mile/Day             & 10.90         & 5.19               & 0.10             & 38.25            \\
Walkability Index of all blocks, 2017                                                & Index                & 4.17          & 2.16               & 0.02             & 12.04            \\
Distance to the nearest metropolitan city (Pop \textgreater 1,000,000)                          & Mile                 & 193.89        & 132.89             & 0.00             & 940.87           \\
Percentage of population travelling more than 30 minutes, 2020                                    & \%                   & 28.09         & 13.14              & 0.00             & 79.01            \\
Number of airports within 50 Kilometers                             & \#                   & 0.92          & 1.12               & 0.00             & 18.00            \\
Percentage of population commuting to work by public transportation, 2020   & \%                   & 0.81          & 1.83               & 0.00                & 32.60     \\ \hline

\multicolumn{6}{l}{\textbf{Health infrastructure and environmental variables}}                                                                                                               \\ \hline
Percentage of population with no health insurance                                              & \%                   & 9.52          & 5.84               & 0.00             & 48.08            \\
Healthcare worker employment density, 2027                                                          & Job/Acre            & 0.12          & 0.13               & 0.00             & 1.36             \\
Number of hospitals per 10k people                                                              & \#            & 0.95          & 1.16               & 0.00             & 8.78             \\
Number of hospital beds per 10k people                                                                  & \#                   & 53.85         & 92.26              & 0.00             & 2338.16          \\
Annual median air quality index (AQI), 2019                                                     & Index                & 35.68         & 12.03              & 0.00             & 131.00 \\ \hline
\end{longtable}

\end{document}